# A Review of Deep Learning Approaches to EEG-Based Classification of Cybersickness in Virtual Reality

Caglar Yildirim
Khoury College of Computer Sciences
Northeastern University
Boston, MA
c.yildirim@northeastern.edu

*Abstract*—Cybersickness is an unpleasant side effect of exposure to a virtual reality (VR) experience and refers to such physiological repercussions as nausea and dizziness triggered in response to VR exposure. Given the debilitating effect of cybersickness on the user experience in VR, academic interest in the automatic detection of cybersickness from physiological measurements has crested in recent years. Electroencephalography (EEG) has been extensively used to capture changes in electrical activity in the brain and to automatically classify cybersickness from brainwaves using a variety of machine learning algorithms. Recent advances in deep learning (DL) algorithms and increasing availability of computational resources for DL have paved the way for a new area of research into the application of DL frameworks to EEG-based detection of cybersickness. Accordingly, this review involved a systematic review of the peer-reviewed papers concerned with the application of DL frameworks to the classification of cybersickness from EEG signals. The relevant literature was identified through exhaustive database searches, and the papers were scrutinized with respect to experimental protocols for data collection, data preprocessing, and DL architectures. The review revealed a limited number of studies in this nascent area of research and showed that the DL frameworks reported in these studies (i.e., DNN, CNN, and RNN) could classify cybersickness with an average accuracy rate of 93%. This review provides a summary of the trends and issues in the application of DL frameworks to the EEG-based detection of cybersickness, with some guidelines for future research.

## I. Introduction

Coeval with the virtual reality (VR) technology itself, cybersickness, or simulator sickness, denotes the aftereffects of exposure to VR, commonly characterized by such physiological symptoms as nausea, dizziness, sweating, and lightheadedness [1 - 3]. While different accounts of the causes of cybersickness are available in the literature [4], sensory mismatch theory is commonly regarded as the most plausible explanation for the occurrence of cybersickness [5]. According to the sensory mismatch theory [6, 7], cybersickness arises from a *mismatch* between visual and vestibular systems while users are exposed to a VR experience. More specifically, the vestibular feedback provided to the user by the vestibular system falls short of accounting for the visual feedback provided by the virtual environment (VE) to produce a virtual locomotion effect. This happens because the vestibular system works on the basis of the vestibular cues provided by the physical environment (not the VE) in response to the position, movement, and orientation of the user in the physical environment. When these vestibular cues are not in accordance with the visual stimuli associated with the position, movement, and orientation of the user in the VE, this leads to a visual-vestibular mismatch, which in turn causes cybersickness [1, 4, 5].

Given the long-standing prevalence of cybersickness in VR, even when using state-of-the-art VR head-mounted displays (HMDs) [3], the measurement of cybersickness has been of particular interest to researchers. The majority of the studies in the literature have exclusively utilized self-reported measures of cybersickness to determine whether users experience cybersickness during a VR experience and to quantify its severity [1, 4]. For instance, one commonly-used self-reported instrument is the Kennedy Simulator Sickness Questionnaire [8], which is now the standard self-reported measure of cybersickness [5]. While these self-reported measures have been shown to yield valid and reliable measurements of cybersickness, they are subjective in nature. Therefore, there has lately been an increasing interest in objectively determining the occurrence of cybersickness based on physiological changes observed during VR exposure [9].

Previous studies have attempted to objectively detect cybersickness from heart rate and heart rate variability [10-12], skin conductance levels [11, 12], and respiration rate [11, 12]. Because of the cognitive and neurovegetative changes associated with cybersickness, brainwaves obtained through electroencephalogram (EEG) have also been used to objectively detect cybersickness [13-19]. Given that such an attempt requires automatic detection of cybersickness on the basis of patterns observed in EEG signals, traditional machine learning algorithms (e.g., Support Vector Machines, Naïve Bayes, and k-Nearest Neighbors) have been extensively applied to the analysis of EEG data for automatic cybersickness detection [20-25]. Moreover, recent advances in deep learning (DL) algorithms, which are aimed at learning representations from data in successive layers of processing in neural networks, and increasing availability of computational resources for DL have paved the way for the application of DL frameworks to EEG-based detection of cybersickness [26-29].

This paper provides a systematic review of the peer-reviewed literature on the application of DL frameworks to the analysis of EEG-data to detect cybersickness during VR exposure. The goals of this review were (a) to compile the relevant studies in this burgeoning area of research, (b) to identify the trends and issues in the use of DL frameworks when classifying EEG-data to detect cybersickness, and (c) to provide methodological guidelines for the design of EEG-based VR experiments for cybersickness detection, preprocessing of data, and selection of DL architectures and hyperparameters.

## II. METHOD

### A. Literature Search

To identify the relevant papers in the literature, repeated searches were conducted on three commonly used databases, namely Web of Science, PubMed, and Google Scholar. Because the application of deep learning to the classification of cybersickness levels using EEG data is rather specific, the original database search was performed to include a wider array of papers, using the following keyword combinations: ((cybersickness OR motion sickness OR simulator sickness) AND (EEG OR electroencephalogram OR electroencephalography OR brainwaves OR brain signals or physiological))

This original search was performed to search through all fields for a given paper and resulted in a total of 408, which was further refined to yield 132 results after filtering and eliminating duplicates. Upon a closer look at the results through their title and abstract, the list of papers was reduced to 33 results.

### B. Eligibility Criteria

In order for a paper to be included in this review, it needed to meet certain eligibility criteria. To begin with, the paper should have attempted to predict cybersickness as the target variable. The paper should also have investigated cybersickness in VR interactions using an HMD. Given the proliferation of modern HMDs, cybersickness is a more prevalent issue for HMDs. Furthermore, EEG patterns observed when participants wear an HMD are expected to be different from those observed when users interact with a CAVE-like immersive environment. Finally, the paper should have collected EEG data and should have used at least one DL architecture.

### C. Screening

The refined list of 33 papers was screened according to the eligibility criteria described in the previous section. Some 13 of these papers were excluded from this review because no deep learning technique was used in the paper. Similarly, nine papers were excluded for not having collected and used EEG data to predict cybersickness and six papers for not having used an HMD. Lastly, one paper was a review paper reporting no empirical data. This screening resulted in a final list of four papers that meet the eligibility criteria [26-29]. Therefore, the current review focused on these four papers.

## III. RESULTS

The final list of four papers [26-29] included in this review were closely scrutinized and encoded to capture a variety of attributes in relation to experimental procedures, data preprocessing steps, and DL architecture choices. Table 1 provides a detailed summary of these papers.

### A. Experimental Design

Experimental design plays a pivotal role when integrating EEG measurements into human-computer interaction experiments in VR. The overall experimental procedure followed by a study directly influences whether DL techniques can be applied to the analysis of the data collected from the experiment. Therefore, in what follows a detailed description of the various experimental design choices made by the previous studies is provided as a practical guide for future studies into this nascent area of research.

#### 1) EEG Devices

Some of the early studies into EEG-based classification of cybersickness levels have used research grade EEG devices [15]. While these EEG devices provide reliable measurements of electrical activity in the brain, they present unique challenges when combined with head-mounted VR headsets in terms of placing both the EEG device and VR headset on users' head at the same time. Therefore, there has been an increasing interest in the use of noninvasive, wireless EEG devices to measure cybersickness while participants wear an HMD [13]. This trend was also observed in the studies included in this review. In fact, with the exception of [26], the remaining three studies used a non-invasive, wireless EEG device to capture brain activity (Emotive Epoc+ or Neurosky MindWave).

Emotiv Epoc+ is a 14-channel wireless EEG device and is perhaps one of the most commonly used mobile EEG solutions available. While the availability of 14 channels is an advantage, the physical shape of the device and how it needs to be fitted on users' head may represent a challenge in future studies. The reason is that if not fitted properly, Emotiv Epoc+ readings will be unreliable. NeuroSky MindWave, on the other hand, is a single-channel EEG device placed on the forehead and collects data from the FP1 position. Admittedly, collecting brain activity data through a single channel has apparent downsides when compared to Emotiv Epoc+. That said, the placement of NeuroSky MindWave headset on users' head is substantially easier and works better with a VR HMD than does Emotiv Epoc+. Considering the fact that these two wireless EEG devices have not been designed with VR integration in mind, there will be some tradeoffs when using these devices in future VR experiments. One potential alternative to these devices is LooxidLink (https://looxidlabs.com/looxidlink/), which is a wireless, 6-channel EEG device designed to be easily attached to modern VR HMDs, such as Oculus Rift and HTC VIVE.

TABLE I. SUMMARY OF REVIEW STUDIES

| Study | Study Attributes | | | | | | | |
|---|---|---|---|---|---|---|---|---|
| | *EEG Device* | *VR Content* | *Data* | *Classification Type* | *Preprocessing* | *Algorithm* | *Hyperparameters* | *Accuracy* |
| [26] | 8-channel 250 Hz & 16 bits | 44 VR videos (16s each) | 30,663,600 samples | Multiclass (Likert scale rating) | Fourier Transform (FFT) | CNN | # Layers: 3<br>Activation: Leaky ReLU<br>Optimizer: Adam<br>Pooling: Max<br>Batch normalization | 87.13% |
| [27] | 14-channel Emotiv Epoc+ | 6 360-degree videos (1-5 min each) | 2,722,269 samples | Binary | Normalization Standardization | DNN | # Layers: 3 (128, 256, 128)<br>Activation: ReLU<br>Output Activation: Sigmoid<br>Input shape: 84<br>Output shape: 32<br>Epochs = 1000<br>Dropout = 0.5 | 98.02% |
| | | | | | | CNN | # Convnet Layers: 3 with (5,1) filter<br>Max Pooling: 1 with (2,1) filter<br>FC Layer: 1 with 100 nodes<br>Activation: ReLU<br>Output Activation: Sigmoid<br>Dropout = 0.5<br>Early stopping | 98.82% |
| [28] | 14-channel Emotiv Epoc+ | 4 360-degree videos (2-3 min each) | 550,000 samples | Binary | Normalization Standardization | DNN | # Layers: 3 (128, 256, 128)<br>Activation: ReLU<br>Output Activation: Sigmoid<br>Input shape: 84<br>Output shape: 32<br>Epochs = 1000<br>Dropout = 0.5<br>Early stopping | 99.12% |
| [29] | NeuroSky Mindwave | 3 360-degree videos (10 min in total) | 78,000 samples | Binary | FFT Power Spectral Density | RNN - LSTM | # Layers: 7 (32, 32, 32, 16, 16, 8, 8)<br>Activation: ReLU<br>Output Activation: Sigmoid<br>Batch size: 100<br>Epochs: 125<br>L1 and L2 regularization<br>Optimizer: RMSProp<br>Time steps: 3 (1, 5, 10 min)<br>Dropout = 0.5 | 1-min step: 83.94%<br>5-min step: 83.33%<br>10-min step: 83.92%" |

*1) VR Headset*

When studying cybersickness in VR, the VR headset used to view and interact with the VE is a key consideration. Based on the small number of studies included in the review, it can be said that various types of VR headsets have been used in prior research, including HTC VIVE, FOVE VR, and smartphone-based VR cardboard. Considering the recent release of untethered, light-weight VR headsets, such as Oculus Quest, we foresee that future research will witness the use of different VR headsets.

*2) VR Environments*

In cybersickness prediction studies, the goal is to ensure that some portion of the participants will experience a bearable amount of cybersickness, in order to differentiate between cybersickness and no-cybersickness classes.

To this end, previous studies have predominantly used 360-degree videos, which participants watched in VR. [26] used 44 short videos depicting an assortment of urban and astrospace scenes, all of which authors noted included visual motion. Similarly, the videos used by [27], [28] and [29] also included visual motion. In line with prior cybersickness research, roller coaster scenes were a popular choice in these previous studies [2-4]. The heavy use of VR content inducing visual motion during the VR experience is compatible with the prior work on the effect of visually-induced motion on cybersickness [13]. Therefore, future studies could utilize the same strategy as well.

One problem with these previous studies, however, is that users were passively exposed to these VR experiences with no explicit interaction between the user and VE. Based on the current evidence from these studies, it is unclear whether we would be able to predict cybersickness levels from EEG data using DL approaches when users are more actively engaged in the VE, as in the case of playing a VR game for instance. Thus, future studies should seek to incorporate different VR experiences with which users can interact to some extent. It should be noted, however, that frequent head and body movements could potentially add more noise to EEG data, which should be taken into account when preprocessing the data for analysis.

*3) Cybersickness measures*

In order to build a cybersickness detection framework, previous studies had users provide labels for the EEG data in the form of self-reported cybersickness levels. For this purpose, these studies used slightly different measures. To begin with, [26] used a multiclass classification approach and asked users to rate their cybersickness levels on a 5-point Likert scale (from 5-extreme sickness to 1-comfortable). The other three studies used a binary classification approach. [27] and [28] asked users to indicate whether they experienced cybersickness during VR exposure, which was dichotomously coded. [29], on the other hand, had users complete the SSQ. The authors then assigned a class label of "sickness" vs. "normal" based on users' score on the SSQ. If the SSQ score was greater than 60, the class label was "sickness", and it was "normal" otherwise.

Considering the problems associated with dividing users into cybersickness vs. no cybersickness classes based on an arbitrary cutoff score on a questionnaire, it might be a better idea to present to users a brief description of cybersickness and its symptoms at the end of the VR experience and to directly ask them to indicate whether or not they experienced cybersickness during their VR exposure. Based on the answer to this binary question, a follow-up question may be displayed, asking users to rate the severity of their cybersickness. Currently, there is a lack of comparative studies into the best approach to encoding the class variable, which is why future studies could build and test different models using these different approaches. Some open questions regarding the labeling of data are as follows: Does a binary classification task lead to better performance results, compared to a multiclass classification task? Is it better to use a dichotomous question or a standard questionnaire, such as SSQ?

*4) Experimental procedures*

All four studies have used similar procedures during data acquisition. In general, experimenters placed the EEG device and VR HMD on users' head, and brain activity was captured during the VR experience. Users then provided self-reported ratings of their cybersickness levels. In the case of multiple VR experiences, users took short breaks to eliminate carry-over effects. One trend among the studies was that users were exposed to multiple VR experiences and provided self-reported assessments of their cybersickness levels for each. This way previous studies have been able to collect larger amounts of data than would be obtained using a single VR experience. For future studies, it would be prudent to devise multiple shorter VR experiences than a standalone long experience, especially when the number of users available for the experiment is limited. Regardless of the chunking of the content, all previous studies [26-29] exposed their participants to VR for a total of 10-15 minutes, which future studies could consider.

*B. Data preprocessing*

Given the sensitivity of EEG data to noise, raw electrical activity signals obtained through an EEG device are typically preprocessed before they are fed into a DL architecture. For studies in which the EEG device used to collect the data did not automatically extract power bands [26, 29], researchers applied Fast Fourier Transform (FFT) to raw EEG data to extract power bands (i.e., alpha, beta, gamma, delta, and theta). Once the power signals were available, the dataset was fed into the architecture without further preprocessing, except for normalization and standardization. [27] and [28] did in fact compare the effect of normalizing and standardizing feature vectors and found that standardized features led to a higher accuracy level than normalized features. One exception to this was [29], in which the authors also applied Power Spectral Density after FFT. As for the input formulation of the preprocessed EEG data into DL models, previous studies have exclusively relied on signal values for the different types of brainwaves, which were fed into the models [26-29].

*C. DL Architectures*

*1) Architecture Design*

The design of DL architectures for a given dataset is experimental in nature. This, coupled with the relatively recent application of DL approaches to VR research, translates into a burgeoning area of research at the intersection of machine learning and VR, which is challenging at the same time, as there are no established guidelines to aid researchers in their selection of various architecture designs.

The four studies included in this review have used deep neural networks (DNN), convolutional neural networks (CNN), and recurrent neural networks (RNN). DNNs are basic neural networks with two or more fully-connected, hidden layers, which are usually represented in a group of layers stacked linearly. CNNs are a special type of DNNs designed to learn local patterns in the data through convolution operations (filtering) in convolutional layers, which are then pooled in pooling layers to reduce the size of the representation and to compute the parameters faster. CNNs are optimized for image processing and commonly used in computer applications. RNNs are another type of DNNs specifically designed to work with sequence data. Unlike DNNs and CNNs, RNNs learn representations from data in an iterative manner, allowing the output of a layer to be used as the input to the same layer in the next time step. RNNs are usually defined by the type of recurrent layers used, with two common layers being Long Short-Term Memory (LSTM) and Gated Recurrent Unit.

Of the five models built in the four papers included in this review, two were DNN-based, two CNN-based, and one RNN-based (LSTM to be more precise). Coming from the same research group, the DNN models reported by [27] and [28] used the same network architecture, with three hidden layers containing 128, 256, and 128 nodes, respectively. The classification accuracy of the DNN model was 98.02% in [27] and 99.12% in [28].

Despite being more commonly applied to image data for computer vision applications, CNNs have been used to analyze the sequence data obtained from the EEG device. The CNN model reported by [26] included three convolutional layers, one pooling layer, and two fully-connected layers. [27]'s CNN model was different in that it included three convolutional layers, one pooling layer, and one fully-connected layer. Of note, [27] converted the EEG signals into a black-and-white image before feeding the data into the network to better utilize the optimized performance of the CNN architecture for the analysis of image data. With this approach, [27] obtained a classification accuracy of 98.82%, whereas the accuracy of [26]'s CNN model was 87.13%.

The RNN model built by [29] was an RNN using LSTM with three time-steps and seven fully-connected hidden layers. The best classification accuracy of the LSTM model was 83.94% with a 60-second step. Given the optimized performance of RNNs on time-series data, the relatively poor performance of the LSTM model compared to the DNN and CNN architecture highlights the need for future research into this area. Due to the stark differences in experimental procedures across these studies (e.g., different EEG devices), no firm conclusions can be drawn in relation to the comparison of the performance of these three DL architectures when classifying cybersickness levels from EEG data. It is possible that the low accuracy rate of [29] can be attributed to the fact that the study used a single-channel wireless EEG devices, whereas an 8-channel scalp EEG device and 14-channel wireless EEG device were used in [26] and [27-28], respectively.

*2) Activation Functions*

Activation functions are crucial hyperparameters of DL architectures that determine the output of nodes and are used to introduce non-linearity to the computation of the output of a node. The most commonly-used activation function for hidden layers in DNN models and convolutional layers in CNN models was the rectified linear unity (ReLU) function, which was used in the four models reported in [27-29]. The activation function used in the CNN model reported by [26] was the leaky ReLU. As for the activation functions used for the output layer, all models used the sigmoid function. While these common choices usually work well for the type of classification tasks used in the prediction of cybersickness from EEG data, future research is warranted to compare different activation functions and identify the optimal combination(s) of these activation functions for cybersickness classification problems.

## IV. DISCUSSION

This systematic review provided an analysis of the existing studies into the application of DL algorithms to the EEG-based detection of cybersickness experienced as a result of exposure to VR. As a result of the systematic search of scientific databases, we found only four papers investigating the EEG-based detection of cybersickness using DL frameworks, which clearly indicates the fact that research into this exciting area is in its infancy. The review indicated that these studies have all been recently published and that they have utilized similar procedures for the design, administration, and analysis of EEG-based VR experiments. In what follows, we provide some guidelines for future research into this area, highlighting trends and issues identified in the studies reviewed here.

*A. Experimental Design*

All four studies used similar experimental procedures in which users were exposed to a VR experience and then provided self-reported assessments of their cybersickness. This way authors were able to construct large datasets involving timeseries EEG data as a 2D tensor of (timestamp x channel) shape for each participant. Based on the trend observed in these studies, future studies should consider using multiple, shorter VR experiences as opposed to a single, longer VR experience to collect more data samples. In so doing, each participant will provide multiple data samples, reducing the number of participants that need to be recruited for future experiments.

Another trend in the literature is to formulate the DL task as a binary classification in which the target variable indicates whether or not users experienced cybersickness during the VR experience. Only one out of the five DL models reported in these four studies have used multiclass classification [26], where users rated the severity of their cybersickness on a Likert scale. The rest of the studies used binary classification and had users make a dichotomous selection (cybersickness vs. not) [27-29]. None of the studies included in this review compared the effect of using binary classification to that of using multiclass classification on the performance of DL models when classifying cybersickness levels. That said, [30] did compare the two when classifying cybersickness based on some other physiological signals and found that binary classification yielded a classification accuracy of 82%, while the same was 56% for ternary classification (no, mild, and severe cybersickness). With multiclass classification, it is more likely to obtain an unbalanced dataset in terms of the distribution of the different categories of the target/class variable, especially when there is a limited number of users participating in the experiment. Therefore, for future studies, it would be prudent to obtain a balanced sample and start out with binary classification. As research into this area flourishes, it is projected that more studies comparing binary classification to multiclass classification will become available.

*B. DL Architectures*

Previous studies included in this review have built DNNs, CNNs, and RNNs to classify cybersickness from EEG

signals. The details of these frameworks are available in Table 2. While these studies clearly explained their models, the descriptions of these models can be improved in several ways, which serve as good guidelines for future studies. To begin with, data preprocessing steps should be clearly outlined in a separate section. This section should provide information on whether the EEG data were transformed to extract power signals or whether raw EEG values were used. If transformations and other feature extraction techniques were applied, these should be clearly explained and justified.

When reporting the structure of DL frameworks, the type of the DL model should be explicitly stated. The description of the DL architecture should clearly state the input formulation (the shape of the feature tensor), the number and type of layers, activation functions used in these layers, and the shape and activation function of the output layer. There should also be a detailed description of other model-specific key hyperparameters, such as filters and pooling layers applied to convolutional layers in CNNs and time steps for RNNs.

Future studies should also elaborately describe how they addressed the problem of overfitting, which occurs when the predictive model is very specific to the training data but cannot generalize to the test data and is even more pronounced for DL frameworks. Activity regularization, dropout, and early stopping are some common strategies to reduce overfitting. They should be used when necessary, and this should be explicitly described. Furthermore, future studies should clearly describe how the data were split into training, validation, and test sets, as well as the evaluation strategy (e.g., k-fold cross validation).

Another issue was that previous studies have solely relied on classification accuracy as a performance metric. Only [29] reported other metrics, such as a confusion matrix. While classification accuracy is extensively reported in the literature, it fails to capture the whole picture when it comes to evaluating the performance of a predictive model. Thus, future studies should report a wider array of classification metrics, such as confusion matrix, F-1 score, AUC-ROC, and log loss. It would also be useful to report training and prediction times for various models to enable other researchers to better assess the tradeoff between the accuracy and speed of these models.

*C. Data and Code Sharing*

One common issue across the studies included in this review was that none of them have made their data or code publicly available. Given the dire need for reproducible AI research in computer science circles [31], it is of paramount importance that studies conducting computational experiments, such as the ones included in this review, publicly share the data and code used for preprocessing and analysis. This simple open-science practice will enable future researchers to devise new data analysis strategies and help broaden the availability of studies into this exciting, but nascent, area of research, while at the same time increasing the credibility of scientific research in this field.